%% file: OmegaPi.tex
\documentclass[pdftex,twocolumn,epjc3]{svjour3}          % twocolum
\RequirePackage[T1]{fontenc}
\smartqed  % flush right qed marks, e.g. at end of proof

\RequirePackage{graphicx}
\RequirePackage{mathptmx}      % use Times fonts if available on your TeX system
\RequirePackage{flushend}
\RequirePackage[numbers,sort&compress]{natbib}
\input{Definitions}
\RequirePackage[colorlinks,citecolor=blue,urlcolor=blue,linkcolor=blue]{hyperref}

\journalname{Eur. Phys. J. C}

\begin{document}

\title{Spin Density Matrix of the $\omega$ in the Reaction \pbarpToOmegaPi0}
\subtitle{Crystal Barrel Collaboration}

\author{
  C.~Amsler\thanksref{addr3, e1} \and
  F.H.~Heinsius\thanksref{addr1} \and
  H.~Koch\thanksref{addr1} \and
  B.~Kopf\thanksref{addr1} \and
  U.~Kurilla\thanksref{addr1, e2} \and
  C.A.~Meyer\thanksref{addr2} \and
  K.~Peters\thanksref{addr1, e2} \and
  J.~Pychy\thanksref{addr1, e3}\and
  M.~Steinke\thanksref{addr1}\and
  U.~Wiedner\thanksref{addr1}
}

\thankstext{e1}{~Now at Albert Einstein Center for Fundamental Physics, Laboratory for High Energy Physics, University of Bern, 3012 Bern, Switzerland}
\thankstext{e2}{~Now at GSI Helmholtzzentrum f\"ur Schwerionenforschung GmbH, 64291 Darmstadt, Germany}
\thankstext{e3}{~This work comprises part of the thesis of J.~Pychy}

\institute{%
~Ruhr-Universit\"at Bochum, 44801 Bochum, Germany\label{addr1}
\and
~Carnegie Mellon University, Pittsburgh, Pennsylvania 15213, USA\label{addr2}
\and
~Physik-Institut der Universit\"at Z\"urich, CH{-}8057 Z\"urich, Switzerland\label{addr3}
}

\date{Received: date / Accepted: date}
% The correct dates will be entered by the editor

\maketitle

%\linenumbers

\begin{abstract}
\input{Abstract}
% \PACS{PACS code1 \and PACS code2 \and more}
% \subclass{MSC code1 \and MSC code2 \and more}
\end{abstract}

%%%%%%%%%%%%%%%%%%%%%%%%%%%%%%%%%%%%%%%%%%%%%%%%%%%%%%%%%%%%%
%% main text
%%%%%%%%%%%%%%%%%%%%%%%%%%%%%%%%%%%%%%%%%%%%%%%%%%%%%%%%%%%%%
\input{Introduction}
\input{CB}
\input{Selection}

\input{PWA}

\input{SDM}

\input{Conclusion}

\end{document}

%% file: Definitions.tex
\def\mevc{MeV/c}
\def\mev2c{\ensuremath{\mathrm{MeV/c^2}}}
\def\gev2c{\ensuremath{\mathrm{GeV/c^2}}}
\def\to{\ensuremath{\,\rightarrow\,}}
\def\pbar{\ensuremath{{\bar{p}}}}
\def\pbarp{\ensuremath{{\bar{p}p}}}
\def\omegapi0{\ensuremath{\omega\pi^0}}

\def\pbarpToOmegaPi0{\ensuremath{{\bar{p}p}\,\rightarrow\,\omega\pi^0}}
\def\OmegaToPi0Gamma{\ensuremath{\omega\,\rightarrow\,\pi^0\gamma}}
\def\OmegaToPipPimPi0{\ensuremath{\omega\,\rightarrow\,\pi^+\pi^-\pi^0}}
\def\PipPimPi0{\ensuremath{\pi^+\pi^-\pi^0}}

\def\lmaxpbarp{\ensuremath{L^{max}_{\pbarp}}}
\def\jpc{\ensuremath{J^{PC}}}

%% file: Abstract.tex
The spin density matrix of the $\omega$ has been determined for the reaction \pbarpToOmegaPi0\ 
with unpolarized in-flight data measured by the Crystal Barrel LEAR experiment at CERN. The two main decay modes of the $\omega$ into 
$\pi^0 \gamma $ and $\pi^+ \pi^-  \pi^0$ have been separately analyzed for various \pbar\ 
momenta between 600 and 1940\,\mevc. The results obtained with the usual method by extracting the matrix elements via 
the $\omega$ 
decay angular distributions and with the more sophisticated method via a full partial wave analysis are in good agreement. A strong spin 
alignment of the $\omega$ is clearly visible in this energy regime and
all individual spin density matrix elements exhibit an oscillatory dependence on
the production angle. In addition, the largest
contributing orbital angular momentum of the \pbarp\ system has been identified for the different beam momenta. It increases from
\lmaxpbarp\,=\,2 at 600\,\mevc\ to \lmaxpbarp\,=\,5 at 1940\,\mevc.

%% file: Introduction.tex
\section{Introduction}
\label{intro_lab}
The spin density matrix of particles originating from \pbarp\ annihilations provides important
information about the underlying production process. The knowledge of this property is
quite scarce in the low energy regime for \pbarp\ in-flight reactions and is, however, very fundamental for
high quality and high statistics future experiments like PANDA~\cite{Lutz:2009ff}. One major physics topic of PANDA
is the spectroscopy of exotic and non-exotic states in the charmonium and open charm mass regions in 
\pbarp\ production or formation processes. 
For the identification of such resonances it is very helpful to know which initial \pbarp\ states contribute
and in particular how the corresponding production mechanism can be described in detail.
The information about the contributing orbital angular momenta
of the initial \pbarp\ system and about the spin alignment of vector
mesons produced in such processes is therefore an excellent key to gain a deeper insight into the production mechanisms. 
Therefore the investigation of the reaction \pbarpToOmegaPi0\ with a relatively simple final state and without complex
decay trees via intermediate resonances provides an excellent access to these questions. The \omegapi0 state
couples only to isospin I\,=\,1 and the C-parity C\,=\,-1 of the \pbarp\ system.

The data presented here have been measured with the Crystal Barrel experiment at LEAR in the years 1995 and 1996.
A partial wave analysis has been performed with the PAWIAN software (Partial Wave Interactive Analysis Software)~\cite{Kopf:2013zz} 
by making use of the helicity formalism and considering the complete reaction chain. Various beam 
momenta have been studied between 600 and 1940\,\mevc\ and for two different $\omega$ decay modes,  
\OmegaToPi0Gamma\ and \OmegaToPipPimPi0\ , respectively. For the neutral
decay mode the polarization of the radiative photon has not been measured
and thus it is needed to average over this property. 

Similar studies of this reaction for the charged decay mode of the $\omega$ have already 
been published in \cite{Anisovich:2002su}. The results presented in the following rely on 
a more accurate data selection and a refined analysis. First preliminary results for the
charged decay mode have already been presented in~\cite{Kopf:2013zz}. 

%% file: CB.tex
\section{Crystal Barrel Experiment}
\label{CB_lab}
The Crystal Barrel detector, which has been described in detail elsewhere~\cite{Aker:1992ny}, 
has been designed with a cylindrical geometry along the beam axis.
The \pbarp\ annihilation took place in a liquid hydrogen target cell with a 
length of 4.4\,cm and a diameter of 1.6\,cm located in the center of the detector. This target was surrounded by a silicon vertex detector. This inner part was surrounded by a jet drift chamber which covered 90\,\% and 64\,\% of 
the full solid angle for the inner and outer layer, respectively. These devices together with a solenoid magnet providing a homogeneous 
1.5\,T magnetic field parallel to the incident beam guaranteed a good vertex reconstruction, tracking and identification for 
charged particles. For accurate measurements of photons the detector was equipped with a barrel of 1380 CsI(Tl) crystals covering the full
azimuthal range of 360$^\circ$ and polar angles from 12$^\circ$ to 168$^\circ$. With this electromagnetic calorimeter, assembled between the jet drift chamber and the solenoid magnet, an energy resolution of 
$\sigma_E/E \approx$ 2.5\,\% and an angular resolution of 1.2$^\circ$ in $\theta$ and $\phi$ each have been achieved.            

%% file: Selection.tex
\section{Data selection and measured angular distributions}
\label{datasel_lab}

The data for this analysis have been taken over various beam times in
the years 1995 and 1996 using an unpolarized \pbar-beam and an unpolarized
liquid hydrogen target. 
In most cases the data samples have been recorded by utilizing a 
0-prong trigger for the neutral and a 2-prong trigger for the charged 
decay mode. In addition, a mixed trigger has been used 
where events with exactly 0 and 2 detected charged particles have been accumulated. 

\begin{figure}[htb]
\centering
\includegraphics[width=0.95\linewidth]{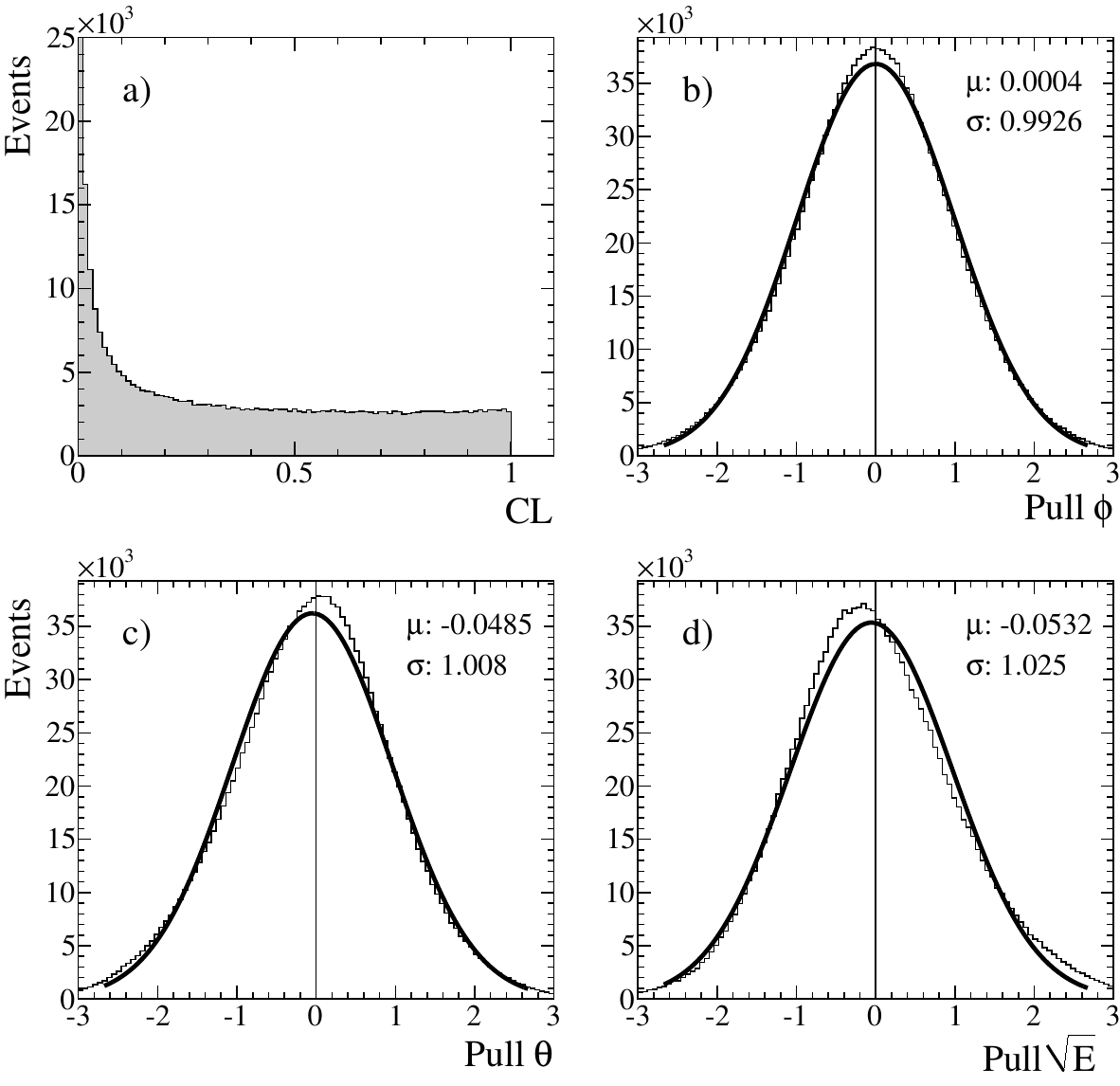}
\caption{Confidence level and pulls resulting from the kinematic fit for the hypothesis \pbarp\to$\pi^0 \pi^0 \gamma$ performed on the all 
neutral events at 900\,\mevc\ beam momentum. The flat distribution of
the confidence level (a) and the parameters of the Gaussian fit (black
lines) to the pull distribution of the angles $\phi$
and $\theta$ (\,(b) and (c)\,) and the square root of the energy (d)
of the reconstructed photons are 
indications for the good data quality and for a well understood error matrix. The
big enhancement at low confidence level values are caused by background and not properly
reconstructed events.}
\label{fig:Pulls}     
\end{figure}

The offline reconstruction and event selection have been performed similarly to the \pbarp\ annihilation at 
rest data~\cite{Amsler:1993kg}. In addition neural networks have been applied for the recognition of misleadingly reconstructed photons 
induced from electromagnetic~\cite{Degener:1995xq} and hadronic~\cite{Berlich:1997qj} split-offs in the ca\-lo\-ri\-meter. 
Only exclusive events are considered where all final state particles have been detected. In order to simply reduce the data samples to a more
manageable size, preselection cuts have been carried out as 
follows: exact number of charged particles and photons in the final state and conservation of the total energy 
($\Delta E^{tot}\,=\,|E^{tot}_{\pbarp}-E^{tot}_{rec}|\,<$\,500\,MeV) and momentum 
($\Delta p^{tot} \,=\,|p^{tot}_{\pbarp}-p^{tot}_{rec}|\, <$\,500\,\mevc) for the desired reaction. In addition exactly one 
$\pi^+\pi^-$ pair must be reconstructed for the charged decay mode originating from a common vertex which is required to be within the 
target cell. After that kinematic fits with the hypotheses \pbarp\to$\, \pi^+ \pi^- 4\gamma$,\-  
$\pi^+ \pi^- 2\pi^0$,\- $\omega 2\gamma$ for the charged decay mode and \pbarp\to$\, 5\gamma$,\- $\pi^0 \pi^0 \gamma$,\- $\omega 2\gamma$  
for the neutral decay mode have been
performed. Each individual fit requires the conservation of the momentum and energy of the events (4 constraints) and additional constraints on
the $\pi^0$-mass. Due to the fact that even with these fits the width of the reconstructed invariant mass of the $\omega$ is still dominated by the 
detector resolution further improvements of the quality of the data has been achieved by constraining the narrow mass of this vector meson (7-constraint fit: 
\pbarp\to$\omega\pi^0$). 
It is required that the fit converges with a 
confidence level (CL) greater than 10\,\% for each hypothesis. For all beam momenta the distribution 
of the confidence level 
is nearly flat and the distributions of the individual
pulls are found to be Gaussian centered at 0 with a 
width of about $\sigma \approx$ 1. This is an indication for a good data quality and for a proper adjusted 
error matrix. As an example Fig.~\ref{fig:Pulls} shows these distributions for the neutral decay mode at 900\,\mevc.          

\subsection{Signal-background separation}
\label{bgr_lab} 
The background contamination is caused by a variety of different
sources. One scenario is that channels decaying to slightly different combinations 
of final state particles contribute where one particle remains undetected or energy 
deposits in the electromagnetic calorimeter
originating from split-off 
effects are misinterpreted as an additional photon. Another
possibility for the fulfillment of all selection criteria is that even channels 
containing the same final state
particles can contribute as background due
to misleadingly combined decay products.

\begin{figure}[htb]
\centering
\includegraphics[width=\linewidth]{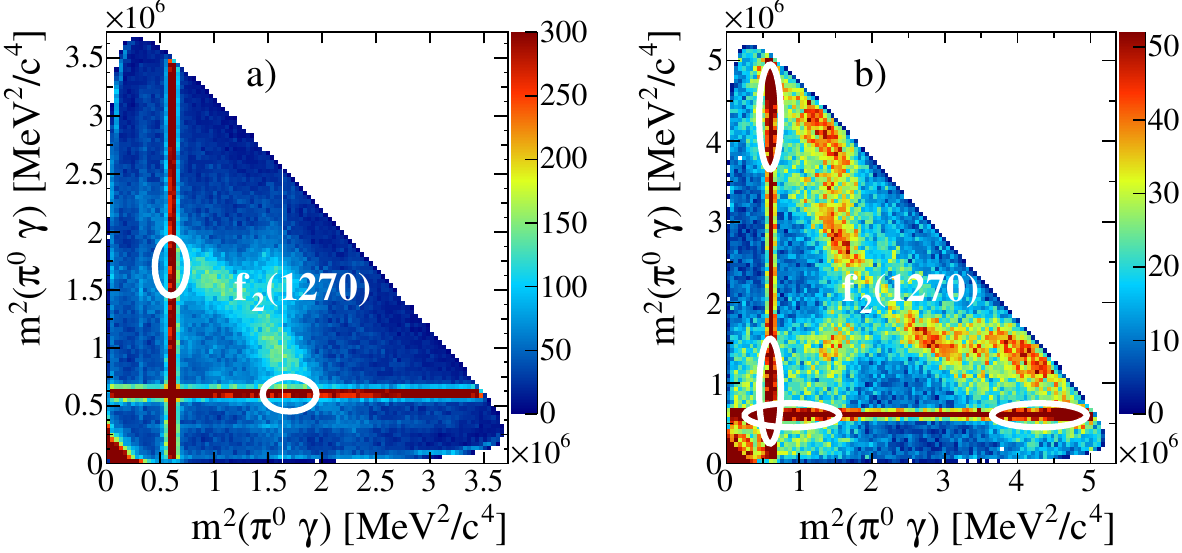}
\caption{Dalitz plots for the selected $\pi^0 \pi^0 \gamma$ events at the 
\pbar\ momentum of 900\,\mevc\ (a) and 1940\,\mevc\ (b). The $\omega$ 
signal is visible as strong narrow bands parallel to the horizontal and 
vertical axis at 
$\approx$ 6$\cdot\mathrm{10^5\,MeV^2/c^4}$. The remaining bands 
mainly originate from the $f_2(1270)\,\pi^0$ background channel. The cross 
regions between the $\omega$ and the
background bands are marked by white ellipses. The comparison between the two 
plots clearly demonstrates that the positions of the crossing regions strongly 
depend on the incident beam momentum.}
\label{fig:DalitzPlots}     
\end{figure}

For the neutral channel the Dalitz plot of the selected $\pi^0 \pi^0
\gamma$ events sheds light on the most crucial background
source (Fig.~\ref{fig:DalitzPlots}). Besides the clear $\omega$ signal, structures from background
events are visible whose major origin has been identified as the channel
\pbarp\to$\mathrm{f_2(1270)}\,\pi^0\,\rightarrow\,(\pi^0\pi^0)\,\pi^0\,\rightarrow\,6\gamma$  
where one photon remains undetected. In this case the most problematic events are those 
which appear in the crossing regions of the signal and background band. 
Due to the fact that in this region the events are located 
in the same phase space volume it is impossible to reject the
background by just applying the selection criteria as described
above. Moreover these inhomogeneities of the background events along the
$\omega$ band whose distribution is directly correlated to the one of
the $\omega$ decay angle would result in huge systematic uncertainties for the
determination of the spin density matrix. Since the positions of 
the crossing regions vary with the incident beam momentum, 
this situation becomes even more problematic.  

In order to separate these non-interfering background sources from the signal events, an elaborated 
technique has been used where a signal weight factor Q has been assigned to each event. The
strategy has been described in detail in~\cite{Williams:2009jinst} and was successfully applied 
on CLAS data for the reaction 
$\gamma\,p\,\rightarrow\,p\,\omega$~\cite{Williams:2009ab, Williams:2009aa}.
Usual separation methods like the side-band subtraction method are based on the requirement 
of a binned data set. This exhibits disadvantages due to the complexity in a high 
dimensional phase space. Instead, the advantages of the technique used here is 
that it is an event based method and that detailed information about the 
specific background sources is not needed. 

\begin{figure}[htb]
\centering
\includegraphics[width=\linewidth]{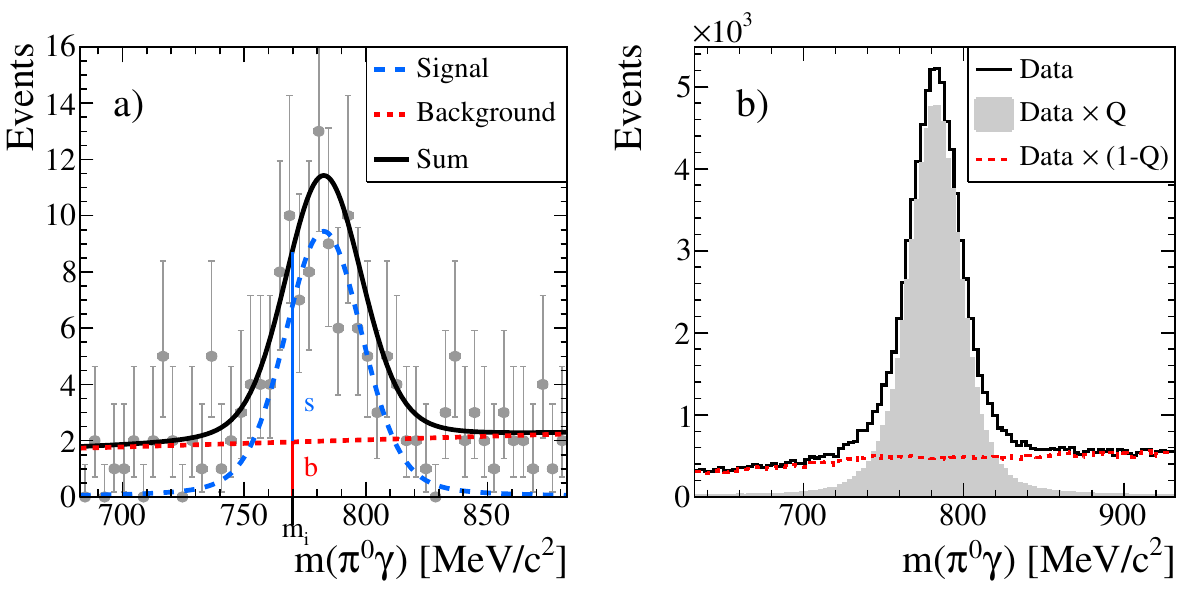}
\caption{ a) Invariant $\pi^0\gamma$ mass from a data subset of the 200 
nearest neighbors associated to a certain $\pi^0\pi^0\gamma$ event at the beam
momentum of 900\,\mevc. The black shape represents the complete fit result 
for the determination of the event weight. The dashed blue line shows the obtained content 
of the $\omega$ signal and the red dotted line the fraction of the background contribution. 
b) Invariant $\pi^0\gamma$ mass of all selected $\pi^0\pi^0\gamma$ events at 900\,\mevc.
The shaded area represents the signal fraction where each event is weighted by its Q-value.
The background content with the individual event weight of (1-Q) is marked with the dotted red line.}
\label{fig:BackgroundResultsNeutral}     
\end{figure}

\begin{figure}[htb]
\centering
\includegraphics[width=\linewidth]{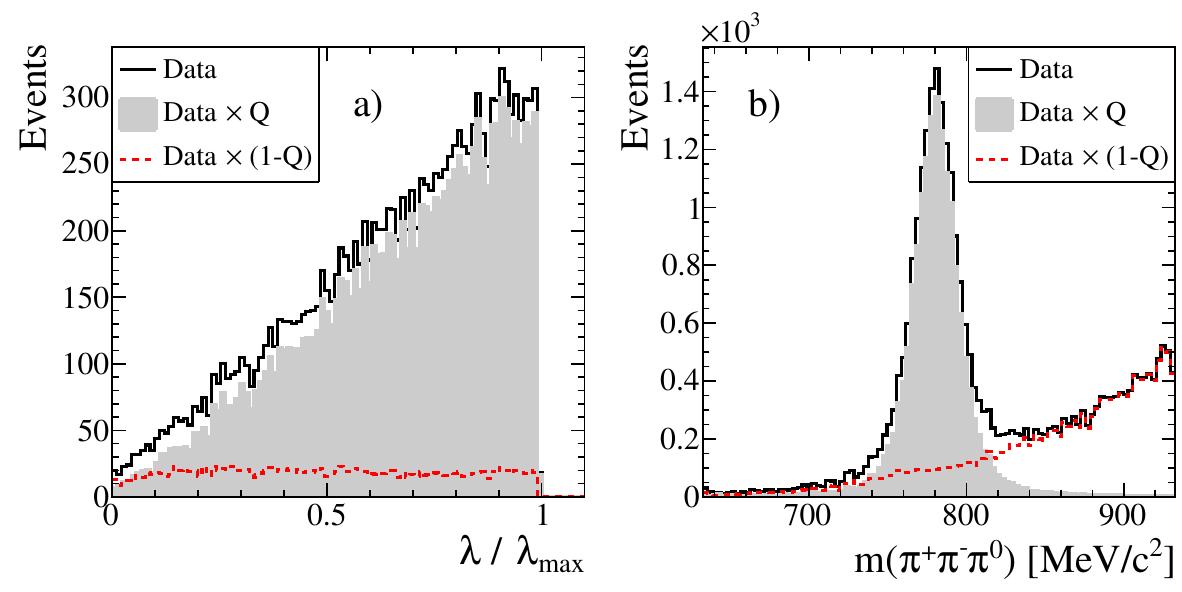}
\caption{ Histogram (a) shows the normalized transition rate $\lambda/\lambda_{max}$ of the $\omega$ decay. 
Histogram (b) represents the invariant $\pi^+\pi^-\pi^0$ mass of all selected 
$\pi^+\pi^-2\pi^0$ events at the beam momentum of 900\,\mevc. The excellent background 
separation power can be seen by the shaded areas representing the fraction of the signal 
events and the dotted red lines illustrating the background content.}
\label{fig:BackgroundResultsCharged}     
\end{figure}

The method takes advantage of the fact that all non-interfering background events cannot reproduce 
the narrow resonance shape of the $\omega$ meson in the corresponding invariant mass spectrum. 
Therefore not the fitted $\omega \pi^0$ events but rather all selected and fitted $\pi^0\pi^0\gamma$ events for 
the neutral and $\pi^+\pi^-\pi^0\pi^0$
events for the charged decay mode appearing within a certain window around the relevant $\omega$-mass
shape (see Fig. \ref{fig:BackgroundResultsNeutral}b, \ref{fig:BackgroundResultsCharged}b)
are considered for the determination of the Q-value.
The procedure starts with the assignment of the 
nearest neighbors for each event by defining a metric with the relevant kinematic observables. 
For the neutral channel the metric has been defined via three observables: the polar 
angle of the $\omega$ production in the \pbarp\ rest frame and the azimuth 
and polar angle of the $\omega$ decay in its helicity system, in which the y-axis is defined to 
be parallel to the normal vector of the production plane. 
A subset of 200 neighbors for each event has been chosen which ensures 
that the associated events cover only a small 
region of the phase space. A Q value for each event is then obtained 
by the determination of the signal to background ratio in the invariant mass spectrum of the 
corresponding data subset. For this an unbinned fit has been performed with a convolution
of a Gaussian and a non-relativistic Breit-Wigner function for the description of the 
$\omega$ signal and a linear approximation for the background content. This approximation
can be justified by the assumption that the background events are homogeneously distributed
within the small region of the phase space. One example of this fit procedure is illustrated in 
Fig.~\ref{fig:BackgroundResultsNeutral}a.  
The invariant $\pi^0\gamma$ 
spectrum (Fig.~\ref{fig:BackgroundResultsNeutral}b) shows the excellent result 
for the global signal-background separation obtained for the beam momentum at 900\,\mevc.  
With the outcome of this approach the final data set for the input of the partial wave analysis has
been selected with the fitted $\omega \pi^0$ events, each weighted with the corresponding Q-factor.

The same event weight method has been performed for the charged decay mode.  
Here, the non-interfering background events exhibit as well different shapes in 
the invariant $\pi^+\pi^-\pi^0$ mass distribution in comparison to the
$\omega$ signal. Potential interfering background sources are channels decaying 
into the same final state particles and can be estimated from
the \pbarp\ annihilation into the four charged pion final state \cite{Bertin:1997vf}, i.e.
\begin{displaymath}
\begin{array}{@{}l}
\pbarp\to\rho^+\rho^-\to(\pi^+\pi^0)\,(\pi^-\pi^0), \\
\pbarp\to\rho^0f_2(1270)\to(\pi^+\pi^-)\,(\pi^0\pi^0), \\
\pbarp\to a_2(1320)^\pm\pi^\mp\to(\rho\pi)^\pm\pi^\mp\to(\pi^\pm\pi^0\pi^0)\,\pi^\mp\quad \mathrm{and}\\ 
\pbarp\to\eta\pi^0\to (\pi^+\pi^-\pi^0)\,\pi^0.
\end{array}
\end{displaymath}
Due to kinematic reasons 
these events do not overlap with the $\omega\pi^0$ events in the phase-space volume 
and thus do not contribute to the background. Also the small fraction of combinatorial background
of these events do not interfere with the $\omega \pi^0$ channel and can therefore be eliminated 
by the Q-weight method. 
For the charged decay mode the metric has been defined with
four independent observables: the polar 
angle of the $\omega$ production in the \pbarp\ rest frame,
the azimuth and polar angle of the normal of the $\omega$ decay plane in its
helicity system and the transition rate $\lambda$ of the $\omega$ decay, which 
is characterized by the cross product of two pion momenta in the $\omega$ 
helicity frame \cite{ Williams:2009ab, Zemach:1964, Weidenauer:1993mv}: 

\begin{eqnarray}
\lambda\,&=&\, |\mathrm{\vec{p}}_{\pi^+}\,\times\,\mathrm{\vec{p}}_{\pi^-}|^2\, /\,\lambda_{max}\\
\mathrm{with}\quad \lambda_{max} &=& T^2\left(\frac{T^2}{108\, c^4} + \frac{m_\pi T}{9\, c^2} + \frac{m_\pi^2}{3}\right),\\
T &=& T_{\pi^+} + T_{\pi^-} + T_{\pi^0},
\end{eqnarray}
where $T_{\pi}$ represents the kinetic energy of the individual pions. 
Figure~\ref{fig:BackgroundResultsCharged}  shows very impressively the obtained background separation power. Especially the shape
of the normalized transition rate $\lambda$ demonstrates the proper distinction between signal and background events. While 
the signal events follow the expected $\lambda$-shape for the $\omega$ decay
with a linear increase and an intersection at the origin of the axis (0 at $\lambda$=0) the
background results in an almost flat distribution.     

\subsection{Overview of the selected data samples}
\label{sumDataSamples}

Tables \ref{tab:DataSel_charged} and \ref{tab:DataSel_neutral}
summarize the numbers of the selected $\omega \pi^0$ events without
and with the obtained weight factor Q for both decay modes. The number 
of $\omega\pi^0$
signal events is found to be between 1\,698 at 1\,525\,\mevc\ and 12\,823 
at 900\,\mevc\ for the charged decay mode and between 1\,113 at 600\,\mevc\
and 53\,788 at 900\,\mevc\ for the neutral decay mode, respectively. The large
variations of the ratio between the event numbers of the two decay modes for the 
different beam momenta are 
mainly caused by the use of different trigger configurations during the data taking.  
The final data sets consist of sufficient numbers of events for achieving significant
results for the partial wave analysis and in particular
for the determination of the spin density matrix of the $\omega$. 
The background contamination estimated by the weight factor (1-Q) depends 
slightly on the beam momentum and on the decay pattern and varies between 
9.2\,\% and 14.6\,\% for the charged and 13.7\,\% and 21.4\,\% for the neutral decay mode.  

\begin{table}[htb]
\caption{Used data samples and number of selected events for the channel $\bar{p} p \rightarrow \omega \pi^0 \rightarrow (\pi^+ \pi^-  \pi^0) \pi^0$.}
\label{tab:DataSel_charged}       % Give a unique label
\footnotesize
\begin{tabular}{r r r r}
\hline\noalign{\smallskip} 
$\bar{p}$ momentum        & total number  &     selected           & signal events  \\ 
  $[$MeV/c$]$       & of events     & $\omega \pi^0$ events  & $\sum Q$       \\
\hline 
  900               & 14\,890\,812  &  14\,460               & 12\,823        \\
  1\,525            & 19\,591\,826  &   1\,871               & 1\,698         \\
  1\,642            & 9\,371\,307   &   3\,475               & 3\,137       \\
  1\,940            & 55\,814\,567  &  10\,942              & 9\,714 \\
\hline
\end{tabular}
\end{table}

\begin{table}[htb]
\caption{Used data samples and number of selected events for the channel $\bar{p} p \rightarrow \omega \pi^0 \rightarrow (\pi^0 \gamma) \pi^0$.}
\label{tab:DataSel_neutral}       % Give a unique label
\footnotesize
\begin{tabular}{r r r r}
\hline\noalign{\smallskip} 
$\bar{p}$ momentum       & total number  &     selected           & signal events  \\ 
  $[$MeV/c$]$       & of events     & $\omega \pi^0$ events  & $\sum Q$       \\
\hline
  600                & 1\,046\,484    &  1\,369               & 1\,113        \\
  900               & 12\,628\,286   &  62\,357               & 53\,788        \\
  1\,050             & 6\,198\,731   &  38\,715              & 33\,236         \\
  1\,350            &  9\,102\,322   &  31\,617               & 25\,933       \\
  1\,525            & 24\,854\,889   &  30\,276               & 24\,980       \\
  1\,642             & 3\,435\,070   &  11\,993               & 9\,926       \\
  1\,800             & 5\,237\,105   &  19\,482               & 15\,763       \\
  1\,940            & 55\,814\,567   &  14\,204                & 11\,169       \\
\hline
\end{tabular}
\end{table}

\subsection{Measured angular and $\lambda$-distributions}
\begin{figure*}[tbp!]
\centering
\includegraphics[width=1.0\textwidth]{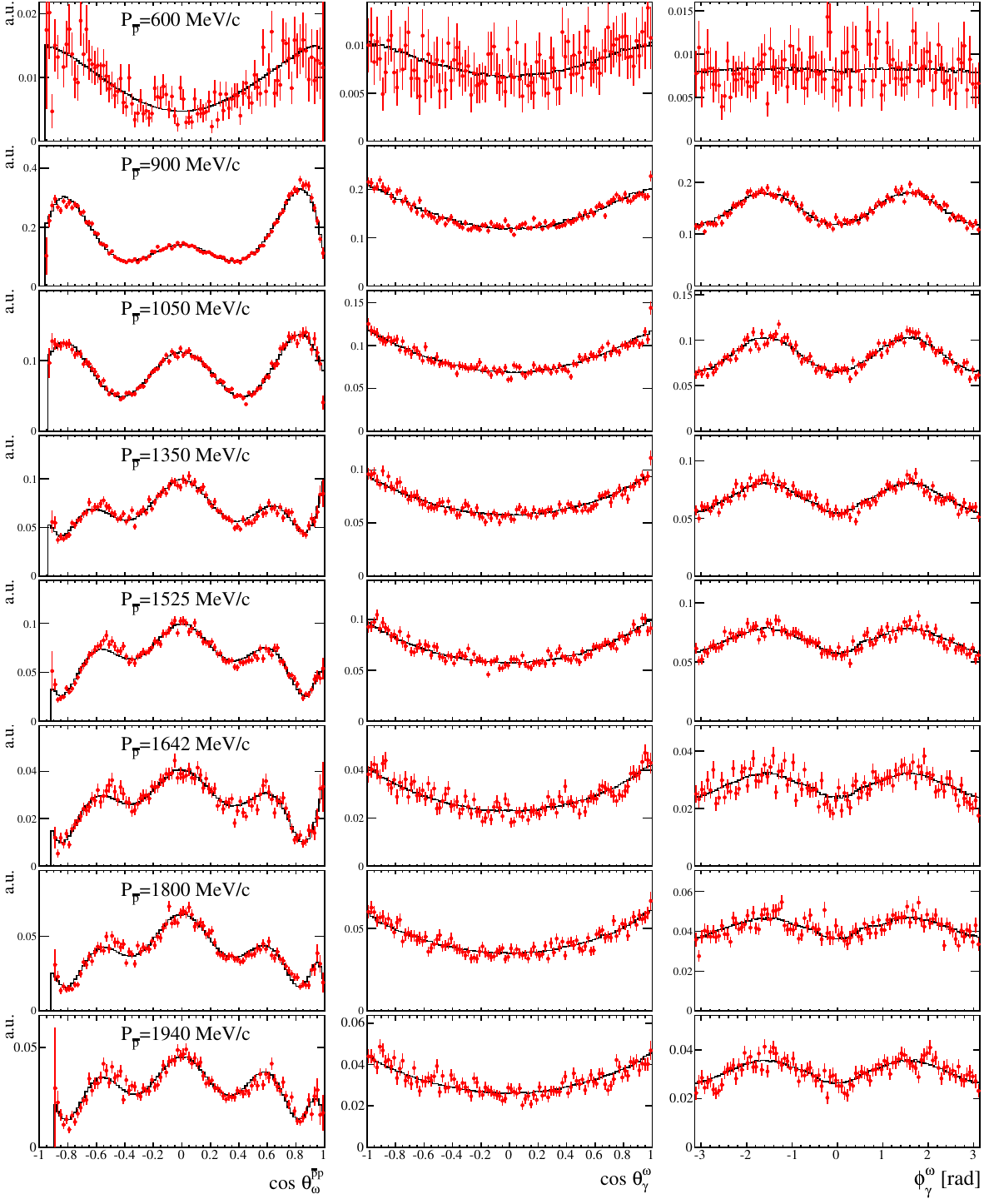}
\caption{Acceptance corrected angular distributions for the channel \pbarpToOmegaPi0\to$(\pi^0\gamma)\pi^0$ as a function of the
production angle (first column) and of the decay angle in cos($\theta^{\omega}_{\gamma}$) (second column) and $\phi^{\omega}_{\gamma}$ (third column). The production angle is defined
in the $\pbarp$ rest frame by the direction of the $\omega$ related to the beam axis. The decay angles are specified by the helicity 
system of the $\omega$ meson. The production angle distribution is given integrated over all
$\omega$-decay angles, the decay angle distributions are given integrated over all
production angles.
While the data are marked with red error bars, the fit results (Sec.~\ref{pwa_lab}) are plotted with black lines. Each row represents 
one specific beam momentum.}
\label{fig:decayAnglesNeutral}
\end{figure*}

\begin{figure*}[htb!]
\centering
\includegraphics[width=1.0\textwidth]{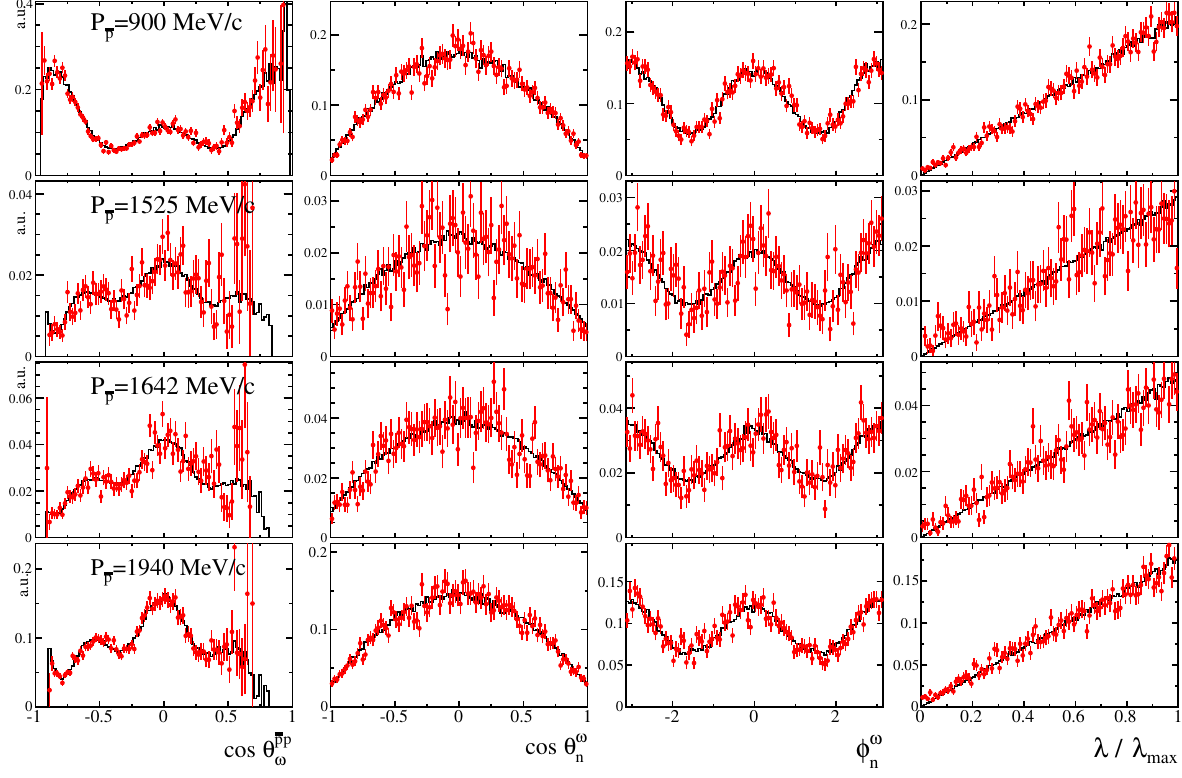}
\caption{Acceptance corrected angular distributions for the channel \pbarpToOmegaPi0\to$(\pi^+\pi^-\pi^0)\pi^0$ as a function of the
production angle  cos($\theta^{\pbarp}_{\omega}$) (first column) and
of the decay angle in cos($\theta^\omega_n$) (second column) and
$\phi^\omega_n$ (third column). The production angle is defined
in the $\pbarp$ rest frame by the direction of the $\omega$ related to
the beam axis. The decay angles $\theta^\omega_n$ and $\phi^\omega_n$ are specified by the normal of the
decay plane of the  $\omega$ meson in its helicity system. The
production angular distribution is integrated over all
$\omega$-decay angles, the decay angular distributions are integrated over all
production angles.
The fourth column represents the normalized transition rate $\lambda$ 
of the $\omega$ decay. A detailed description of this property can be found in sec.~\ref{bgr_lab}.
While the data are marked with red error bars, the fit results (Sec.~\ref{pwa_lab}) are plotted with black lines. Each row represents one 
specific beam momentum.}
\label{fig:decayAnglesCharged}
\end{figure*}

Fig.~\ref{fig:decayAnglesNeutral} and
Fig.~\ref{fig:decayAnglesCharged} show the relevant angular distributions 
obtained from the \omegapi0 data after applying all selection and
background rejection criteria for the neutral and charged decay mode, respectively. The 
distributions of the $\omega$ production angle are integrated over the $\omega$-decay
distributions and are characterized by fluctuations of the intensity 
with a higher number of extrema for increased beam momenta. This is an indication that 
more waves contribute with the rise of the center of mass energy. The huge error bars and 
the absence of entries around $\left| \, \cos(\theta^{\pbarp}_\omega)\,\right|$~=~1 are caused by the 
acceptance leakage of the detector in the very forward and backward region. These inefficiencies
are more distinctive for the charged decay mode due to the limited angular coverage
of the tracking devices and become even more apparent with increasing beam momentum. 
The distributions of the $\omega$-decay angles are integrated over all production angles and
exhibit typical shapes for this particle (see Sec. \ref{sdm_lab}).
 
For all beam momenta the normalized
$\lambda$-distributions for the $\omega$-decay to $\pi^+\pi^-\pi^0$
(Fig.~\ref{fig:decayAnglesCharged}) 
are in excellent agreement with the expected shape. This illustrates again the high purity of
the \omegapi0 data for the individual beam momenta.  

%% file: PWA.tex
\section{Partial wave analysis}
\label{pwa_lab}
\subsection{Amplitudes}
\pbarp\ in-flight reactions where mesons and photons 
are exclusively involved are dominated by the s-channel process. Therefore the  
partial wave analyses for those reactions have been 
started usually with the \jpc\ system initiated from the \pbarp\ annihilation. One
difficulty of this method is that additional Clebsch-Gordan coefficients 
for the coupling of the \pbarp\ system with the \jpc\ intermediate
state are not considered correctly. In order to avoid such error-prone 
procedure, the analysis performed on the data here is based on the 
description of the complete reaction chain starting from the \pbarp\ 
coupling up to the final states. This new method is summarized in
detail in~\cite{HKoch:2012} and can also be applied to
other \pbarp\ reactions in flight.   

The starting point is the description of the differential cross section
of the whole reaction chain where the transition 
amplitude depending on the helicities of the involved particles 
is divided into the $\omega$-production and the $\omega$-decay amplitude. 
For the neutral channel this cross section is expressed by

\begin{eqnarray}
\label{eq:crossSection}
\frac{d\sigma}{d\tau} \propto w \, = \, 
\sum_{\scriptsize
\begin{array}{@{}l}\lambda_{\bar{p}}, \lambda_p,\lambda_{\pi^0_r}(=0), \\\lambda_{\pi^0_d}(=0), \lambda_\gamma
\end{array}
}
\Big|\sum_{\lambda_{\omega}} & T_{\lambda_{\bar{p}} \lambda_p
  \lambda_{\pi^0_r} \lambda_{\omega}}^{\pbarp \rightarrow
  \omega\pi^0_r} (cos(\theta^\pbarp_\omega)) \nonumber \\
 & \cdot  A_{\lambda_{\omega} \lambda_{\pi^0_d}
   \lambda_\gamma}^{\omega\rightarrow\pi^0_d\gamma}
 (cos(\theta^\omega_\gamma), \phi^\omega_\gamma) \Big|^2 , 
\end{eqnarray}
where $d\tau$ represents the infinitesimal volume element of the phase-space, 
$w$ the transition probability, $\lambda$ the helicities of all involved particles, 
$T$ the production and $A$ the decay amplitude in the helicity frame. The two neutral
pions are distinguished by the notation $\pi^0_r$ for the recoil 
particle and $\pi^0_d$ for the $\omega$ decay particle. Due to the fact that
a mass constraint for the $\omega$ has been used for the kinematic fit the dynamics for 
this meson (e.g. a Breit-Wigner distribution) has not been taken into account.
It is noteworthy to mention that
the components of the transition amplitude are added coherently
over the helicities of the intermediate $\omega$-
resonance and incoherently over the helicities of all initial and
final state particles. Eq.~\ref{eq:crossSection} is expanded into states with
definite $J^{PC}$-values defining the partial wave helicity amplitudes
$T^{J^{PC}}_{\lambda_{\pbar}, \lambda_p, 0, \lambda_\omega}$ and
$A^{1^{--}}_{\lambda_\omega \lambda_\gamma}$. These partial wave amplitudes
are further expanded in states with definite $J^{PC}$, $L$,
$S$-values where $L,S$ are the respective orbital angular momenta
and total spins of the \pbarp, \omegapi0 and $\pi^0\gamma$-system
($L_{\pbarp}$, $S_{\pbarp}$, $L_{\omegapi0}$, $S_{\omegapi0}$(=1),
$L_{\pi^0\gamma}$(=1), $ S_{\pi^0\gamma}$(=1)), defining the amplitudes \linebreak 
$T_{L_{\pbarp}, S_{\pbarp}, L_{\omegapi0}}^{J_{\pbarp}}$ and
$A_{L_{\pi^0\gamma}, S_{\pi^0\gamma}}^{1} = A^1_{11}$.
Here, the 
quantum number $J$ represents the total angular momentum, $L$ the orbital angular momentum and $S$ the total spin
of the related system composed of two particles. The underlying 
formalism for theses expansions can be found in
detail elsewhere~\cite{Chung:1971ri}. With the requirement that the
parity, charge conjugation and total angular momentum
are conserved for strong and electromagnetic interactions
the differential cross section can be
described by incoherent sums over the \pbarp\ singlet and
triplet states and over the helicity of the radiative photon of the
final state system \cite{HKoch:2012}. 
In terms of $LS$-amplitudes Eq.~\ref{eq:crossSection}
reads \cite{HKoch:2012}:
\begin{eqnarray}
\label{eq:finalProdAmpPbarP}
w & = & \sum_{\lambda_\gamma, \lambda_p,\, \lambda_{\pbar}} 
\Big|  \sum_{J_\pbarp} \sum_{L_\pbarp, S_\pbarp} \sum_{L_{\omegapi0}, \lambda_\omega}
\, \sqrt{2L_\pbarp+1}  \nonumber \\
&& \cdot  \langle L_\pbarp, 0 , S_\pbarp, M_\pbarp | J_{\pbarp}, M_{\pbarp}\rangle \,  
\langle \frac{1}{2}, \lambda_\pbar , \frac{1}{2}, -\lambda_p | S_{\pbarp},  M_{\pbarp} \rangle \nonumber \\
&& \cdot \sqrt{2L_{\omegapi0}+1}  \, \langle L_{\omegapi0}, 0 , 1,
\lambda_\omega | J_{\pbarp},  \lambda_\omega  \rangle 
\, d^{J_{\pbarp}}_{M_{\pbarp} \, \lambda_\omega}(\theta^{\pbarp}_\omega) \nonumber \\
&& \cdot 
T_{L_{\pbarp}, S_{\pbarp}, L_{\omegapi0}}^{J_{\pbarp}}  \, \cdot \, \sqrt{\frac{3}{8\pi}}\nonumber \\ 
&& \cdot
\,
D^{1*}_{\lambda_\omega\lambda_\gamma}
(\theta^\omega_\gamma,\phi^\omega_\gamma) \cdot
A_{11}^1 \cdot \lambda_\gamma  
\Big|^2,
\end{eqnarray} 
with $M_{\pbarp}=\lambda_\pbar-\lambda_p$. The summation over $\lambda_\pbar$ and $\lambda_p$ 
can be arranged in such a way, that one incoherent term
for singlet states ($S_{\pbarp}=0, M_{\pbarp}=0$) and three incoherent
terms for triplet states  ($S_{\pbarp}=1, M_{\pbarp}=0,\pm1$) appear.  
The direction of the \pbar\ beam is chosen as the quantization axis 
which results in the restriction of the z-component of $J_{\pbarp}$ to $M_{\pbarp}=0,\pm1$. 
The \omegapi0\ system is fully characterized by  $L_{\omegapi0}$, $S_{\omegapi0}$=1, the helicity 
$\lambda_\omega$ and the production angle $\theta_\omega^{\pbarp}$ of the $\omega$ 
in the \pbarp\ rest frame. Due to the fact that  the \pbarp\ system is unpolarized the angle 
$\phi_\omega^{\pbarp}$ is not defined. 
The $\omega$ decay system is characterized by the 
angular momentum $L_{\pi^0_d\gamma}$=1, the total spin $S_{\pi^0_d\gamma}$=1, the helicity $\lambda_\gamma$
and the decay angles $\theta_\gamma^\omega$ and $\phi_\gamma^\omega$ of the $\gamma$ in the helicity frame 
of the $\omega$ meson.  The $d_{M_\pbarp,\lambda_\omega}^{J_\pbarp}$ denotes the
Wigner-d function for the decay of the \pbarp\ system, 
$D_{\lambda_\omega,\lambda_\gamma}^{*J_\omega=1}$ the complex conjugate of the Wigner-D function
for the $\omega$ decay and 
$\langle L,0, S, \lambda_1-\lambda_2 | J, \lambda_1-\lambda_2  \rangle$
and
$\langle j_1,\lambda_1, j_2,- \lambda_2 | S, \lambda_1-\lambda_2
\rangle$
the Clebsch-Gordan coefficients for the $LS$- and $jj$-coupling
respectively. As $A^{1}_{11}$ for a given \pbarp-energy is a fixed complex number, 
the product $T_{L_{\pbarp}, S_{\pbarp}, L_{\omegapi0}}^{J_{\pbarp}}
\cdot A^{1}_{11}$  is handled as one complex 
parameter $\alpha^{J^{PC}}_{L_{\pbarp} L_{\omega{\pi^0}}}$.

For the reaction \pbarpToOmegaPi0\,$\rightarrow\,(\pi^+\pi^-\pi^0) \,
\pi^0$ Eq.~\ref{eq:finalProdAmpPbarP} has to be modified \cite{Chung:1971ri}.
The incoherent sum over $\lambda_\gamma$ vanishes and 
the $\omega$ decay amplitude $A^{1^{--}}_{\lambda_\omega \lambda_\gamma}$ 
has to be replaced by

\begin{equation}
\sqrt{\frac{3}{4\pi}}\cdot D^{1*}_{\lambda_\omega\mu}
(\Theta^\omega_n,\Phi^\omega_n, \gamma^\omega_n) \cdot A^1_{\mu}(E_{\pi^+}, E_{\pi^-}) ,
\end{equation}
where $\Theta^\omega_n,\Phi^\omega_n, \gamma^\omega_n$ are the Euler angles of the
normal of the $3\pi$-decay plane ($\vec{n}$) in the $\omega$-helicity system
with $\mu = \langle \vec{J}_\omega\cdot\vec{n}\rangle$. In general $\mu$ takes the
values $\pm 1, 0$, but in the $\omega\rightarrow\pi^+\pi^-\pi^0$ case,
only $\mu = 0$ is allowed. $A^1_{\mu}(E_{\pi^+}, E_{\pi^-})$ describes the amplitude
in the Dalitz plot, which is proportional to $|\vec{P}_{\pi^{+}} \times \vec{P}_{\pi^{-}}|$
\cite{Zemach:1964}.

By making use of the conservation principles and the selection rules one can 
easily extract the specific combinations of the relevant quantum numbers 
allowed for the reaction \pbarpToOmegaPi0\ (Tab.~\ref{tab:QuantenNo}).

\begin{table}[htb]
\caption{Combinations of the allowed quantum numbers for the reaction \pbarpToOmegaPi0. The \jpc\ 
combinations even$^{+-}$ and odd$^{-+}$ are forbidden for the \pbarp\ system 
and the combinations even$^{-+}$, even$^{++}$ and odd$^{++}$ are not allowed for the
\omegapi0\ coupling. The quantum numbers for the $\omega$ decay to $\pi^0\gamma$ 
(\,$L_{\pi^0\gamma}$=1, $S_{\pi^0\gamma}$=1\,) and to $\pi^+\pi^-\pi^0$ (\,$L_{\pi^+\pi^-\pi^0}$=1,
 $S_{\pi^+\pi^-\pi^0}$=0\,) and $S_{\omegapi0}$=1 for the \omegapi0\ coupling are 
not given explicitly.}

\label{tab:QuantenNo}      
\footnotesize
\centering
\begin{tabular}{l c c c c}
\hline\noalign{\smallskip} 
  \jpc       & $L_\pbarp$  & $S_\pbarp$  &    $M_\pbarp$     & $L_{\omegapi0}$   \\ 
\hline
  0$^{--}$    &           \multicolumn{4}{c}{not allowed for \pbarp\ reaction}       \\
  even$^{--}$ &  J         &  1          &   $\pm$1       &  J-1, J+1        \\
  odd$^{--}$  &  J-1, J+1  &  1          &  0 ,$\pm$1       &  J       \\
  odd$^{+-}$  &  J         &  0          &   0                & J-1, J+1 \\
\hline
\end{tabular}
\end{table}

\subsection{Fits to data and determination of the parameters $\alpha$}
\label{pwa_lab_fit}
Unbinned maximum likelihood fits were performed for each beam momentum and decay mode
individually in order to determine the best hypothesis 
with the resulting fit parameters $\alpha^{\jpc}_{L_{\pbarp} L_{\omegapi0}}$. Input for
this method are the selected data with the obtained event weights $Q_i$ as well as
phase-space distributed Monte Carlo events. For properly taking into account the detector 
resolution and acceptance the GEANT3 transport code has been used.
To considering also the correct reconstruction efficiency these Monte Carlo
events were then undergoing the same reconstruction and selection criteria as applied for data events
and described in section~\ref{datasel_lab}. 
The extended likelihood
function $\mathcal{L}$ is defined as~\cite{SBischoff:1999}:
\begin{eqnarray}
\label{equ:likelihood}
\mathcal{L} \propto n_{data}!\cdot\exp\Big(-\frac{(n_{data}-\overline{n})^2}{2n_{data}}\Big)\cdot\prod_{i=1}^{n_{data}} \frac{w(\vec{\tau_i}, \vec{\alpha}) \, \epsilon(\vec{\tau_i})}
{\int w(\vec{\tau}, \vec{\alpha}) \, \epsilon(\vec{\tau}) \, \mathrm{d}\tau} \nonumber & \\
\end{eqnarray}
where $n_{data}$ denotes the number of data events, 
 $\vec{\tau}$ the phase-space coordinates, $\vec{\alpha}$ the complex fit parameter, 
$\epsilon(\vec{\tau})$ the acceptance and reconstruction efficiency at the position $\vec{\tau}$ and
$\overline{n}=n_{data} \cdot \int w(\vec{\tau}, \vec{\alpha}) \, \epsilon(\vec{\tau}) \, \mathrm{d}\tau / 
\int \epsilon(\tau)\,\mathrm{d}\tau$.
The $w(\vec{\tau}, \vec{\alpha})$ represents the transition probability given by Eq.~\ref{eq:finalProdAmpPbarP}.
By logarithmizing Eq.~\ref{equ:likelihood}, approximating the integrals 
with Monte Carlo events and introducing the weight $Q_i$ for each event, the final 
function to be minimized is then given by:
\begin{eqnarray}
\label{equ:likelihoodAprox}
 -\ln\,\mathcal{L} &\approx& -\sum_{i=1}^{n_{data}} \ln (w(\vec{\tau_i}, \vec{\alpha})\cdot Q_i) \nonumber \\
       & &   +\Big(\sum_{i=1}^{n_{data}} Q_i\Big) \, \cdot \, \ln\Big( \frac{\sum_{j=1}^{n_{MC}} w(\vec{\tau_j}, \vec{\alpha})}{n_{MC}} \Big) \nonumber \\
       & & + \frac{1}{2} \, \cdot \, \Big(\sum_{i=1}^{n_{data}} Q_i\Big) \, \cdot \, \Big(\frac{\sum_{j=1}^{n_{MC}} w(\vec{\tau_j}, \vec{\alpha})}{n_{MC}}-1\Big)^2 , 
\end{eqnarray}
where $n_{MC}$ represents the number of selected Monte Carlo events.

To obtain the best hypothesis for the description of the data a
strategy has been carried out where fits with successive increase of 
the maximal contributing orbital angular momentum $L^{max}_{\pbarp}$ have been 
performed. For each of those fits all allowed waves with
$L_{\pbarp} \leq L^{max}_{\pbarp}$ have been taken into account. The fit
results have been compared using the likelihood ratio.
With this strategy it was feasible to determine unambiguously the best 
hypothesis and thus the largest contributing orbital angular momentum $L^{max}_{\pbarp}$
for all data samples. Summaries of the obtained
results are listed in 
Tab.~\ref{tab:FitResultCharged} and  Tab.~\ref{tab:FitResultNeutral},
 respectively. Except for the beam momentum of 1525\,\mevc\ the results for the 
charged and neutral decay modes are consistent. The slight discrepancy for
only one beam momentum is likely caused
by the limited acceptance of the detector for the charged decay mode and thus the results for
the neutral decay mode are more reliable.  For the initial states
$J^{PC} = even^{--}$ and $J^{PC} = odd^{+-}$ two different orbital
angular momenta $L_{\omegapi0} = J-1$ and  $L_{\omegapi0} = J+1$ for
the  $\omegapi0$-system are possible (see Tab. \ref{tab:QuantenNo}) . It turned out that for all
fits both waves for this system contribute. As an example the fit result with the obtained parameter 
values are summarized in Tab.~\ref{tab:ParameterExample} for the charged
decay mode at the beam momentum of 900\,\mevc.  

\begin{figure}[htb]
\centering
\includegraphics[width=0.33\textwidth, height=0.27\textwidth]{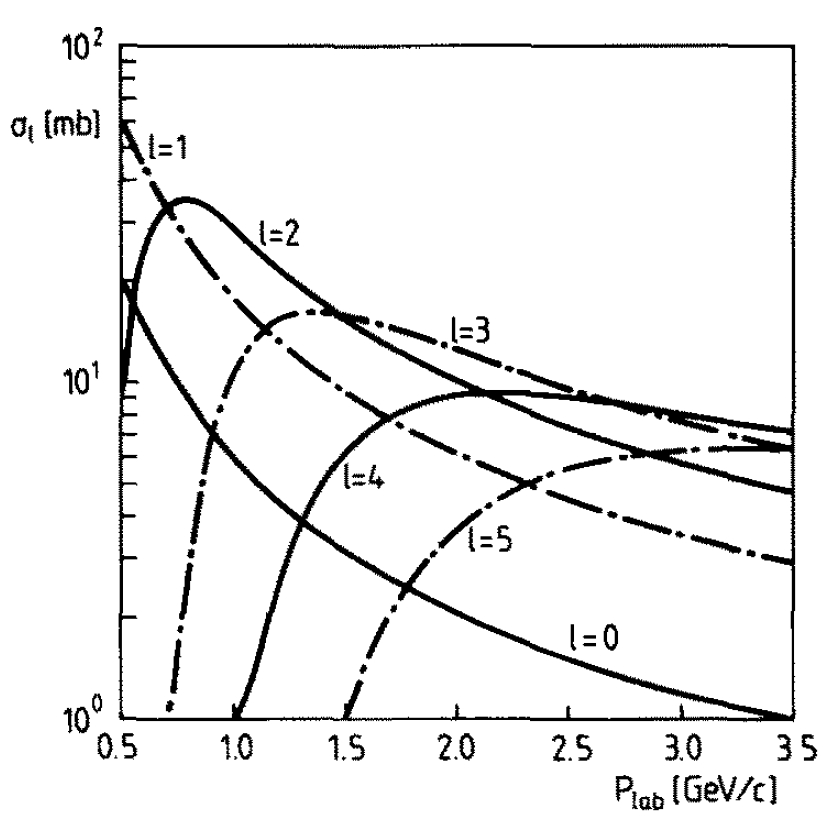}
\caption{Estimated partial wave annihilation cross sections as a function of the 
\pbar\ beam momentum for several $L_\pbarp$ based on model calculations~\cite{Mundigl:1991jp,Weise:1993py}.
The main input parameter for this model is the nucleon radius, which is assumed to be 
$\langle r_B^2 \rangle^{1/2}\, = \,0.6 \,fm$. This figure is extracted from~\cite{Mundigl:1991jp}.}
\label{fig:cs_pbarp_model}     
\end{figure}

The maximal contributing orbital angular momentum \- $L_\pbarp^{max}$ increases continuously 
from 2 at the lowest beam momentum of 600\,\mevc\ up to 5 at the highest
beam momentum of 1940\,\mevc. These values are in good agreement with
a former analysis~\cite{Abele:2000qq}. Partial wave annihilation cross
sections as a function of the \pbar\ beam momentum for several
$L_\pbarp$ values have been estimated \cite{Mundigl:1991jp,Weise:1993py}.
Figure~\ref{fig:cs_pbarp_model} shows the outcome of these model
calculations for a typical hadronic radius of the baryon core of $\langle r_B^2 \rangle^{1/2}\, = \,0.6 \,fm$. 
Under this assumption the minimum \pbar-beam momentum for the production of
$L_\pbarp = 3$ states is expected to be roughly 0.7\,GeV, for $L_\pbarp = 4,5$ states
it is expected to be 1.0 and 1.5\, GeV, respectively.
The results presented here are in good agreement
with these model calculations and only differ in slightly lower momentum
thresholds.

\begin{table}[htb]
\centering
\caption{Best fit results for $L^{max}_{\pbarp}$ for the channel
  \pbarpToOmegaPi0\to(\PipPimPi0)\,$\pi^0$. The significant
  improvement in comparison to the hypothesis with $L_\pbarp^{max}-1$ 
  and the marginal improvement of the assumption with $L_\pbarp^{max}
  +1$ is a good indication for the unambiguousness of the fit result. 
  The significance is denoted in units of the standard deviation $\sigma$.}
\label{tab:FitResultCharged}      
\footnotesize
\begin{tabular}{r c l l }
\hline\noalign{\smallskip} 
  momentum       & $L_\pbarp^{max}$  &
  \multicolumn{2}{c}{significance of likelihood ratio}     \\ 
  $[$\mevc$]$          &                       
  &  $\frac{\ln L(L_\pbarp^{max})}{\ln L(L_\pbarp^{max}-1)}$  &  $\frac{\ln L(L_\pbarp^{max} +1)}{\ln L(L_\pbarp^{max})}$  \\ \hline
  900   &  4  &   \quad 2.2\,$\sigma$  & \quad 0.13\,$\sigma$  \\
  1525  &  4  &   \quad 9.0\,$\sigma$  & \quad 0.90\,$\sigma$ \\
  1642  &  5  &   \quad 3.2\,$\sigma$   & \quad 0.06\,$\sigma$ \\
  1940  &  5  &   \quad >10\,$\sigma$  & \quad 1.04\,$\sigma$  \\
\hline
\end{tabular}
\end{table}

\begin{table}[htb]
\centering
\caption{Best fit results for $L_\pbarp^{max}$ for the channel
  \pbarpToOmegaPi0\to$(\pi^0\gamma)\pi^0$. For further explanations see
  the caption of Tab.~\ref{tab:FitResultCharged}.}
\label{tab:FitResultNeutral}      
\footnotesize
\begin{tabular}{r c l l}
\hline\noalign{\smallskip} 
  momentum       & $L_\pbarp^{max}$  &
  \multicolumn{2}{c}{significance of likelihood ratio}     \\ 
  $[$\mevc$]$          &                       
  &  $\frac{\ln L(L_\pbarp^{max})}{\ln L(L_\pbarp^{max}-1)}$  &  $\frac{\ln L(L_\pbarp^{max} +1)}{\ln L(L_\pbarp^{max})}$  \\ \hline
  600  &  2   &   \quad >10\,$\sigma$   & \quad 1.05\,$\sigma$  \\
  900  &  4   &   \quad 6.5\,$\sigma$   & \quad 0.22\,$\sigma$  \\
 1050  &  4   &   \quad >10\,$\sigma$   & \quad 0.01\,$\sigma$ \\
 1350  &  5   &   \quad 5.6\,$\sigma$   & \quad 0.03\,$\sigma$  \\
 1525  &  5   &   \quad >10\,$\sigma$   & \quad 0.25\,$\sigma$ \\
 1642  &  5   &   \quad 5.0\,$\sigma$   & \quad 8$\cdot$10$^{-3}\,\sigma$ \\
 1800  &  5   &   \quad >10\,$\sigma$   & \quad 0.55\,$\sigma$ \\
 1940  &  5   &   \quad >10\,$\sigma$   & \quad 0.69\,$\sigma$  \\
\hline
\end{tabular}
\end{table}

\begin{table}[htb]
\centering
\caption{Fitted parameters for the reaction \pbarpToOmegaPi0\,$\rightarrow\,(\pi^+\pi^-\pi^0)\,
\pi^0$ at
900 MeV/c beam momentum and $L^{max}_{\pbarp}=4$. The L,S correspond to the
$\overline{p}p$, $\omega\pi^{0}$ or $\omega$ system, respectively.
For technical reasons
the $\alpha^{\jpc}_{L_{\pbarp} L_{\omegapi0}}$ were split into two parameters where
some can be fixed due to the linear dependency. Furthermore, two
additional phases can be fixed as only relative phases are relevant.
The errors are statistical only and originated from the 
covariance matrix obtained by the fit.}
\label{tab:ParameterExample}      
\footnotesize
\begin{tabular}{l c c}
\hline\noalign{\smallskip} 
Parameter for $J^{PC} \, (L,S)$ & Magnitude & Phase    \\ 
\hline
$\overline{p}p\ \mathrm{system}$ & & \\
$1^{--} \, (0,1)$ & 0.4 $\pm$ 0.04 & 0 (fixed) \\
$1^{--} \, (2,1)$ & 0.13 $\pm$ 0.08 & 1.8 $\pm$ 0.4 \\
$1^{+-} \, (1,0)$ & 0.36 $\pm$ 0.03 & 0 (fixed) \\
$2^{--} \, (2,1)$ & 0.20 $\pm$ 0.04 &-2.86 $\pm$ 0.19 \\
$3^{--} \, (2,1)$ & 0.16 $\pm$ 0.06 &1.49 $\pm$ 0.27 \\
$3^{--} \, (4,1)$ & 0.15 $\pm$ 0.06 &1.47 $\pm$ 0.24 \\
$3^{+-} \, (3,0)$ & 0.190 $\pm$ 0.028 &2.09 $\pm$ 0.19 \\
$4^{--} \, (4,1)$ & 0.135 $\pm$ 0.016 &0.9 $\pm$ 0.3 \\
$5^{--} \, (4,1)$ & 0.08 $\pm$ 0.04 &1.9 $\pm$ 0.5 \\
$\omega\pi^{0}$ production & & \\
$1^{--} \, (1,1)$ & 1 (fixed) & 0 (fixed) \\
$1^{+-} \, (0,1)$ & 0.70710 (fixed) &0  (fixed) \\
$1^{+-} \, (2,1)$ & 0.62 $\pm$ 0.15 &1.20 $\pm$ 0.17 \\
$2^{--} \, (1,1)$ & 0.70710 (fixed) &0  (fixed) \\
$2^{--} \, (3,1)$ & 0.85 $\pm$ 0.14 &0.71 $\pm$ 0.28 \\
$3^{--} \, (3,1)$ & 1 (fixed) &0  (fixed) \\
$3^{+-} \, (2,1)$ & 0.70710 (fixed) &0 (fixed) \\
$3^{+-} \, (4,1)$ & 0.85 $\pm$ 0.11 &0.22 $\pm$ 0.24 \\
$4^{--} \, (3,1)$ & 0.70710 (fixed) &0  (fixed) \\
$4^{--} \, (5,1)$ & 0.8 $\pm$ 0.1 &0.69 $\pm$ 0.16 \\
$5^{--} \, (5,1)$ & 1 (fixed) &0  (fixed) \\
$\omega$ decay & &\\
$1^{--} \, (1,0)$ & 1(fixed) & 0 (fixed)\\
\hline
\end{tabular}
\end{table}

\subsection{Comparison of data and fits}
The fitted $\omega$-production and $\omega$-decay angles and the
normalized $\lambda$-value
(in case of the charged $\omega$-decay) are compared with the data in 
Fig.~\ref{fig:decayAnglesNeutral} and Fig.~\ref{fig:decayAnglesCharged}.
Apart from minor systematic discrepancies the agreement is good. The reasonable description of the data can also be seen in the fit quality 
summarized in Tab.~\ref{tab:DataQuality}. The goodness-of-fit has been estimated with
the Pearson $\chi^2$ test based on the histograms for the relevant kinematic 
variables (Fig.~\ref{fig:decayAnglesNeutral} and Fig.~\ref{fig:decayAnglesCharged}) by calculating

\begin{eqnarray}
\label{equ:pearsonTest}
\frac{\chi^2}{ndf}  = \sum_{i=1}^{n} \sum_{j=1}^{N_{bins,i}} \Big(\frac{(\nu_{ij,fit}-\nu_{ij,data})^2}{\nu_{ij,data}} \Big)
/ (N_{bins}-N_{params}),
\end{eqnarray} 
where n represents the number of the relevant kinematic variables, $N_{bins,i}$ the number of bins for the histogram $i$,\linebreak
$\nu_{ij,data/fit}$ the number of data/fit entries within bin $j$ for the histogram $i$ and $N_{params}$ the 
number of fit parameters. The $\chi^2$ values divided by the number of degrees of 
freedom vary between 0.82 and 1.36 which are reasonable results. However, one has to remark that Eq. \ref{equ:pearsonTest} does 
not consider the correlations between the different kinematic variables and thus only serves as a rough estimate for the fit quality.

\begin{table}[htb]
\centering
\caption{Fit quality $\chi^2/ndf$ for the charged and neutral decay mode obtained with the Pearson $\chi^2$ test.}
\label{tab:DataQuality}      
\footnotesize
\begin{tabular}{r c c}
\hline\noalign{\smallskip} 
  momentum       & \multicolumn{2}{c}{$\chi^2/ndf \; (ndf)$}     \\ 
  $[$\mevc$]$      &  $\omega$\to$\pi^+\pi^-\pi^0$  & $\omega$\to$\pi^0\gamma$ \\ \hline
  600  &  --    &  0.82 (282)   \\
  900  &  1.16 (371)   &  1.36 (274)  \\
 1050  &  --    &  1.18 (273)   \\
 1350  &  --    &  1.04 (268)   \\
 1525  &  1.13 (356)   &  1.13 (268)   \\
 1642  &  1.04 (352)   &  1.27 (267)   \\
 1800  &  --    &  1.21 (267)   \\
 1940  &  1.02 (351)   &  1.20 (267)   \\
\hline
\end{tabular}
\end{table}

%% file: SDM.tex
\section{Spin density matrix of the $\omega$}
\label{sdm_lab}

In addition to the contributing orbital momenta, the polarization observables of the 
$\omega$ meson exhibit important information about its production process. 
These properties are in general defined by spherical 
momentum tensors or alternatively  by the spin density matrix $\rho$, which is used 
in the following.
Since the $\omega$ is a particle with spin 1 its spin density matrix contains 3x3
complex elements $\rho_{\lambda_i\lambda_j}$, where $\lambda_i$ and $\lambda_j$ 
represent the helicities of the $\omega$-particle.
The $\rho$-matrix is hermitian with a trace of 1 by definition. Polarization 
means $\rho_{11} \neq \rho_{-1-1}$ and 
alignment is defined as 
$\rho_{11} = \rho_{-1-1} \neq \rho_{00}$.  For measurements with unpolarized protons
and antiprotons for
channels where the parity is conserved and by choosing the quantization
axis to be in the production plane, the number 
of independent $\rho$-elements is reduced to four real quantities. The $\omega$ spin density matrix 
for the reaction \pbarpToOmegaPi0\ is given by~\cite{Schilling:1969um}:
\begin{eqnarray}
\label{eq:rhoMatrix}
\rho = \left(\begin{array}{ccc} 1/2(1-\rho_{00}) & \Re\rho_{10}+i\Im\rho_{10} & \rho_{1-1} \\
        \Re\rho_{10}-i\Im\rho_{10} & \rho_{00}  & -\Re\rho_{10}+i\Im\rho_{10} \\
	 \rho_{1-1}  & -\Re\rho_{10}-i\Im\rho_{10} & 1/2(1-\rho_{00})
	 \end{array}\right)
\nonumber \\
\end{eqnarray}
The $\rho$-matrix elements are dependent on the quantization axis which is here
chosen to be the one of the $\omega$-helicity system defined by the $\omega$ flight 
direction in the \pbarp\ center of mass system. The helicity system is the most suitable one to use for this
kind of \pbarp\ reactions which is strongly dominated by the s-channel process. In
addition, the elements are dependent on the center of mass energy and
on the production angle. 

\begin{figure*}[htp!]
\centering
\includegraphics[width=1.0\textwidth]{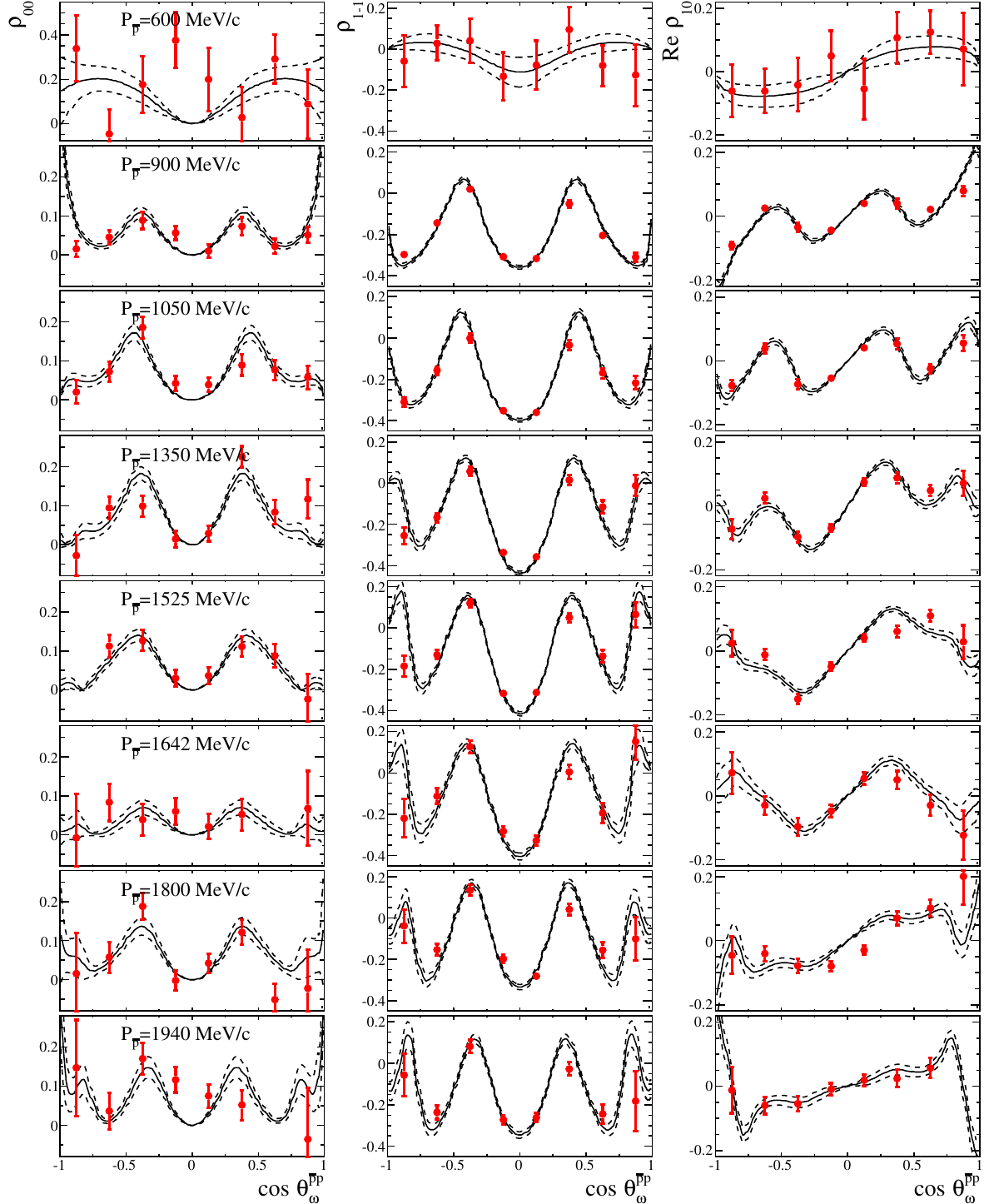}
\caption{Spin density matrix elements $\rho_{00}$ (first column),   $\rho_{1-1}$ (second column) and
 $\rho_{10}$ (third column) of the $\omega$ in its helicity frame as function of the production angle 
for the reaction \pbarpToOmegaPi0\to$(\pi^0\gamma)\pi^0$.
While the results obtained via the $\omega$-decay angles are marked with red error bars, the outcome 
via the partial wave analysis
is plotted with continuous black lines. The dashed black lines represent the statistical errors of the
partial wave result. Each row represents one specific beam momentum. }
\label{fig:sdmNeutral}
\end{figure*}

\begin{figure*}[htb!]
\centering
\includegraphics[width=1.0\textwidth]{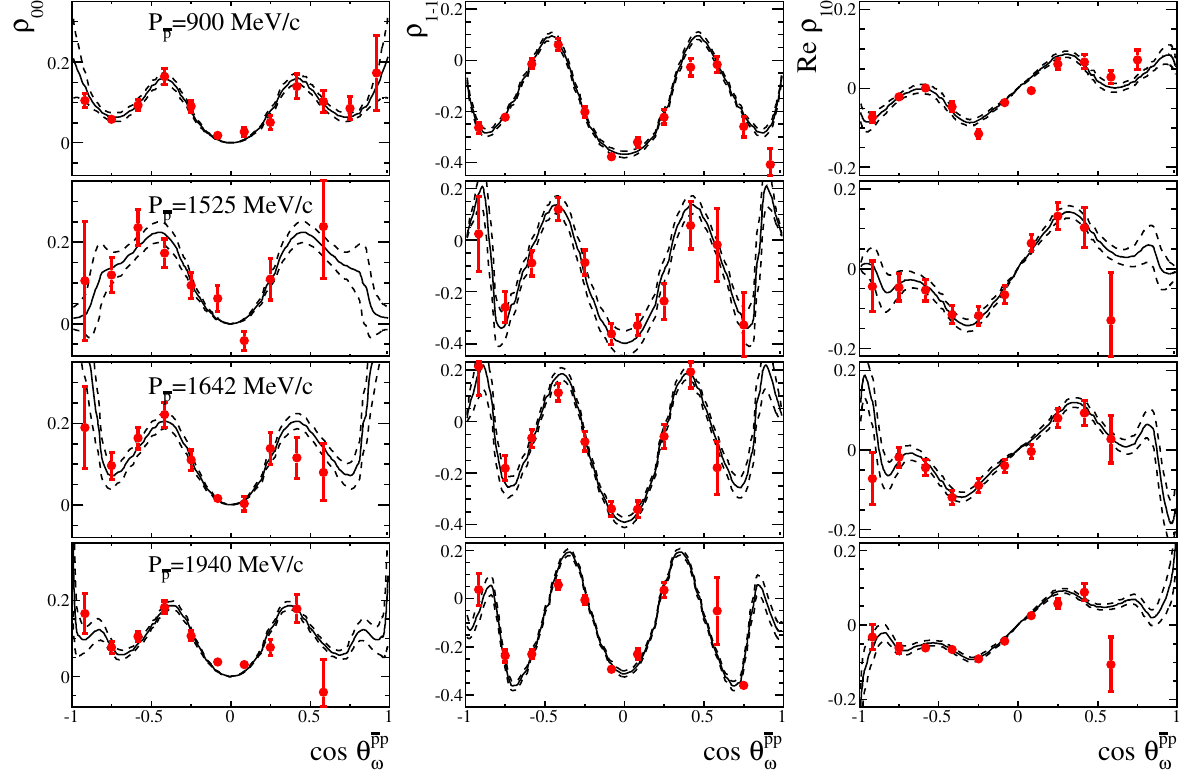}
\caption{Spin density matrix elements $\rho_{00}$ (first column),   $\rho_{1-1}$ (second column) and
 $\rho_{10}$ (third column) of the $\omega$ in its helicity frame as function of the production angle 
for the reaction \pbarpToOmegaPi0\to$(\pi^+\pi^-\pi^0)\pi^0$.
While the results obtained via the $\omega$-decay angles are marked with red error bars, the outcome 
via the partial wave analysis is plotted with continuous black lines. The dashed black lines represent the statistical errors of the
partial wave result. Each row represents one specific beam momentum.}
\label{fig:sdmCharged}
\end{figure*}

The determination of the $\omega$-matrix elements has been performed by 
two different methods: (1) by using the results of the partial wave analysis
and (2) solely via the angular decay distributions of the $\omega$-meson.
The first method is very rarely used and has already
been applied successfully for the reaction
$\gamma\mathrm{p}\,\rightarrow\,\mathrm{p}\omega$
~\cite{Williams:2009aa}. It uses the fitted production amplitude, here defined as 
$T_{\lambda_{\bar{p}} \lambda_p \lambda_{\pi^0_r} \lambda_{\omega}} (\pbarpToOmegaPi0)$ 
(Sec. \ref{pwa_lab}), which contains the information of the $\omega$ 
spin density matrix. The individual $\rho$-elements can be extracted 
from the production amplitude by~\cite{Kutschke:1996}:
\begin{eqnarray}
\label{eq:sdmPwa}
\rho_{\lambda_i\lambda_j}= \frac{1}{N}\sum_{\lambda_{\pbar},\lambda_{p}, \lambda_{\pi^0_r}=0} 
T_{\lambda_{\bar{p}} \lambda_p \lambda_{\pi^0_r} \lambda_i}^*T_{\lambda_{\bar{p}} \lambda_p \lambda_{\pi^0_r} \lambda_j}, 
\end{eqnarray}
where $N$ is the normalization factor:
\begin{eqnarray}
\label{eq:sdmPwaNorm}
N =\sum_{\lambda_{\pbar},\lambda_{p},\lambda_{\omega}, \lambda_{\pi^0_r}=0} 
|T_{\lambda_{\bar{p}} \lambda_p \lambda_{\pi^0_r} \lambda_{\omega}}|^2 
\end{eqnarray}
According to Eq.~\ref{eq:sdmPwa} the $\rho$-matrix elements have been
projected out from the production amplitude obtained from the partial
wave fit of the full reaction chain. 
In our case (unpolarized initial states) the differential cross section is only
dependent on $\rho_{00}$, $\rho_{1-1}$ and $\Re \rho_{10}$, so that only these matrix elements can be extracted.
The results as function of the center of mass
energy and the $\omega$-production angle are summarized in Fig.~\ref{fig:sdmNeutral} 
and \ref{fig:sdmCharged}.
The statistical errors have been calculated by propagating the covariance matrix obtained
by the likelihood fit. Additionally, a much more time consuming bootstrap approach
as described in \cite{Vernarsky:2014}
has been tested which yielded results that are in full agreement to the first calculation.

The second and more traditional method, also called\linebreak
Schilling method, does not make use 
of the results of the partial wave analysis 
and uses only the distribution of the $\omega$ decay angles $\theta$ and
$\phi$~\cite{Schilling:1969um}. The angular distribution for the charged decay mode is given by:
\begin{eqnarray}
\label{eq:schillingOmTo3Pi}
W(\theta^{\omega}_{n} , \phi^{\omega}_{n}) &=& \frac{3}{4\pi} \sum_{\lambda_\omega\lambda_\omega^\prime} \, D^{*1}_{\lambda_\omega 0}(\theta^{\omega}_{n}, \phi^{\omega}_{n})
                                     \, \rho_{\lambda_\omega\lambda_\omega^\prime} \, D^{1}_{\lambda_\omega^\prime 0}(\theta^{\omega}_{n}, \phi^{\omega}_{n})
                       \nonumber \\
                    &=& \frac{3}{4\pi} \, \Big( \frac{1}{2} \, (1-\rho_{00}) \, + \, \frac{1}{2} \, (3\rho_{00}-1) \, \cos^2\theta^{\omega}_{n} 
                       \nonumber \\ 
                    &&  -\sqrt{2} \, \Re\rho_{10} \, \sin2\theta^{\omega}_{n} \, \cos\phi^{\omega}_{n} \nonumber \\ 
		    &&   - \rho_{1-1} \, \sin^2\theta^{\omega}_{n} \, \cos2\phi^{\omega}_{n} \Big),
\end{eqnarray}
and for the neutral decay mode by:
\begin{eqnarray}
\label{eq:schillingOmToPiGamma}
W(\theta^{\omega}_{\gamma}, \phi^{\omega}_{\gamma}) &=& \frac{3}{4\pi} \sum_{\lambda_\omega\lambda_\omega^\prime \lambda_\gamma} D^{*1}_{\lambda_\omega \lambda_\gamma}(\theta^{\omega}_{\gamma},\phi^{\omega}_{\gamma})
                                     \rho_{\lambda_\omega\lambda_\omega^\prime} D^{1}_{\lambda_\omega^\prime \lambda_\gamma}(\theta^{\omega}_{\gamma}, \phi^{\omega}_{\gamma})
                       \nonumber \\
                    &=& \frac{3}{4\pi} \Big( \frac{1}{2} \, (1-\rho_{00}) \, + \, \frac{1}{2} \, (1-3\rho_{00}) \, \cos^2\theta^{\omega}_{\gamma} 
                       \nonumber \\ 
                    &&  + \, \sqrt{2} \, \Re\rho_{10} \, \sin2\theta^{\omega}_{\gamma} \, \cos\phi^{\omega}_{\gamma} \nonumber \\ 
		    &&  + \, \rho_{1-1} \, \sin^2\theta^{\omega}_{\gamma} \, \cos2\phi^{\omega}_{\gamma} \Big)
\end{eqnarray}
As can be seen from Eq.~\ref{eq:schillingOmTo3Pi} and \ref{eq:schillingOmToPiGamma} only the elements of the real part 
of the matrix are sensitive to the $\omega$ decay angular distribution, which are $\rho_{00}$, $\rho_{1-1}$ and $\Re\rho_{10}$. 
The imaginary part $\Im\rho_{10}$ related to an eventual $\omega$-polarization
perpendicular to the scattering plane is not accessible. The matrix elements have been extracted separately 
for different bins in the production angle by 
fitting the two dimensional $\omega$ decay distribution
according to Eq.~\ref{eq:schillingOmTo3Pi} for the charged decay mode and Eq.~\ref{eq:schillingOmToPiGamma} for the neutral 
decay mode 
with a maximum likelihood fit procedure analogous to the one described before in sec. \ref{pwa_lab_fit}. While the two methods rely on different approaches, 
both, however, should yield 
the same results. Due to the fact that binned data in the production angle are needed for the Schilling method the 
determination of the $\rho$-matrix elements with this method is not as accurate as for the first one which naturally imposes all the 
physical constraints and correlations. 

The good agreement between the results obtained with the two different methods can clearly be 
seen in Fig.~\ref{fig:sdmNeutral} for the neutral decay mode and in Fig.~\ref{fig:sdmCharged} 
for the charged decay mode. 
It is noticeable that in case of the PWA method the statistical errors are smaller in comparison
to the Schilling method.
The $\rho$-matrix elements show a strong oscillatory dependence on the
$\omega$-production angle cos($\theta^\pbarp_\omega$). The $\rho_{00}$ and $\rho_{1-1}$ have minima and maxima
for $\cos(\theta^\pbarp_\omega) = 0$ and $|\cos(\theta^\pbarp_\omega)| = 0.4$, respectively.
The $\rho_{00}$-values averaged over the production
angle are listed in Tab. \ref{tab:AveragedRho}. These values show a clear spin alignment effect ($\rho_{00} = 1/3$ would
correspond to no spin-alignment).

\begin{table}[htb]
\caption{$\rho_{00}$-values of the $\omega$ meson averaged over the
  production angle. The averaging is 
limited on the range of the production angle 
with a reasonable detector acceptance, which is between $-0.85 \leq \cos \theta^\pbarp_\omega \leq 0.4$ for the charged and 
$-0.85 \leq \cos \theta^\pbarp_\omega \leq 0.95$ for the neutral decay mode. Only the statistical errors are 
listed below. The systematic errors are not considered here.}
\label{tab:AveragedRho}
\footnotesize
\centering
\begin{tabular}{rcc}
\hline\noalign{\smallskip}
momentum    & \multicolumn{2}{c}{$\overline{\rho_{00}}$} \\
$[$\mevc$]$ & $\omega\rightarrow\pi^+\pi^-\pi^0$ & $\omega\rightarrow\pi^0\gamma$ \\
            & ($-0.85 \leq \cos \theta^\omega_n \leq 0.4$) & ($-0.85 \leq \cos \theta^\omega_\gamma \leq 0.95$) \\
\hline
600  & -                 & 0.15 $\pm$ 0.05\\
900  & 0.069 $\pm$ 0.008 & 0.047 $\pm$ 0.008\\
1050 & -                 & 0.064 $\pm$ 0.011\\
1350 & -                 & 0.075 $\pm$ 0.012\\
1525 & 0.106 $\pm$ 0.016 & 0.065 $\pm$ 0.009 \\
1642 & 0.094 $\pm$ 0.013 & 0.028 $\pm$ 0.012\\
1800 & -                 & 0.060 $\pm$ 0.013\\
1940 & 0.083 $\pm$ 0.007 & 0.060 $\pm$ 0.015\\
\hline
\end{tabular}
\end{table}

The results for the charged and the corresponding neutral decay mode are in an 
overall good agreement for all beam momenta. However, differences are 
visible which are in particular strongly depending on the production angle. These 
inconsistencies are more significant for the results obtained with the PWA method 
due to the relatively small statistical errors and might be caused by systematic 
uncertainties in the simulation and reconstruction procedure.

Similar dependencies on the production angle have already been observed for 
the tensor polarisation observables of the $\omega$ in the same reaction 
\pbarpToOmegaPi0 \cite{Anisovich:2002su}. 
In addition the values obtained in the analysis here
can be compared with earlier vector meson production 
experiments in $\pbarp$-interactions at higher energies. Also there
explicit alignment effects for the $\rho$-meson have been observed \cite{Lednicky:1983}.
This is in contrast to $pp$-reactions, where negligible
alignment for $\rho^0$ is reported \cite{Blobel:1973wr}. This trend is also observed in low
energy pp-reactions for the orientation of the $\omega$-spin \cite{AbdelBary:2010er}.

%% file: Conclusion.tex
\section{Summary}
\label{concl_lab}
The reaction \pbarpToOmegaPi0\ with unpolarized in-flight data 
has been analyzed in detail. The $\omega$ meson with the neutral 
decay to $\pi^0\gamma$ as well as with the charged decay to 
$\pi^+ \pi^-  \pi^0$ has been investigated separately in the low
energy regime for various \pbar\ beam momenta between 600 and 
1940\,\mevc. An excellent background rejection power has been 
achieved by determining an event based signal weight factor. The 
performed partial wave analysis has taken into account the
complete reaction chain starting from the \pbarp\ coupling up
to the final state particles. It described the data with high
precision. The maximal contributing orbital angular momentum 
$L^{max}_\pbarp$ increases continuously from
2 at the lowest beam momentum of 600\,\mevc\ up to 5 at the 
highest beam momentum of 1940\,\mevc. The elements of the spin
density matrix have been determined with two different methods.
The results based on the outcome of the partial wave analysis and those 
based on the $\omega$
decay distributions are in excellent agreement. The first method
via the production amplitudes of the PWA  was only used in a few cases 
up to now.
The individual elements exhibit a strong dependency on the $\omega$-production
angle. A clear spin alignment with $\rho_{00}$ values between
0\% and 25\% over the whole angular range  within 
$|\cos(\theta)|\,<\,$0.9 is visible.